# Generative AI Usage of University Students: Navigating Between Education and Business

**Research Paper**

Fabian Walke[1], Veronika Föller[2]

[1] Heilbronn University of Applied Sciences, Faculty of Business, Heilbronn, Germany
fabian.walke@hs-heilbronn.de
[2] University of Hagen, Faculty of Business Administration and Economics, Hagen, Germany
veronika.foeller@studium.fernuni-hagen.de

**Abstract.** This study investigates generative artificial intelligence (GenAI) usage of university students who study alongside their professional career. Previous literature has paid little attention to part-time students and the intersectional use of GenAI between education and business. This study examines with a grounded theory approach the characteristics of part-time students' GenAI usage. Eleven students from a distance learning university were interviewed. Three causal and four intervening conditions, as well as strategies were identified, to influence students' use of GenAI. The study highlights both the potential and challenges of GenAI usage in education and business. While GenAI can significantly enhance productivity and learning outcomes, concerns about ethical implications, reliability, and the risk of academic misconduct persist. The developed grounded model offers a comprehensive understanding of GenAI usage among students, providing valuable insights for educators, policymakers, and developers of GenAI tools seeking to bridge the gap between education and business.

## 1 Introduction

Generative Artificial Intelligence (GenAI) has become a crucial component of modern technology, as evidenced by the widespread popularity of ChatGPT (OpenAI, 2025). ChatGPT has experienced a remarkable growth in its user base, reaching 100 million users in just two months following its release in November 2022 (Gordon, 2023). The usage of ChatGPT and other GenAI tools has induced a series of disruptive effects across a broad spectrum of human activities (Saúde *et al.*, 2024). GenAI has also gained significant popularity among students, leading to extensive debates about the role of GenAI applications in higher education (Wecks *et al.*, 2024). This trend is reflected in the substantial increase in scientific research since 2023 addressing the impact of GenAI usage in higher education (Saúde *et al.*, 2024). GenAI improves learning outcomes, productivity, and student engagement by opening up new opportunities for individualized education, feedback, and assistance (Adiguzel *et al.*, 2023). However, there are also drawbacks, such as the absence of human interaction, the limited comprehension of GenAI models, and inherent biases in the training data (Baidoo-Anu & Owusu Ansah, 2023). Furthermore, academic integrity is at risk when GenAI tools are misused by students and researchers to generate texts, scientific reports, or even entire theses, presenting these AI-generated works as their own original contributions (Chan, 2023). Additionally, other studies have shown that the use of GenAI tools can negatively affect exam performance (Wecks *et al.*, 2024).

A key area of interest within the broad field of current research is the adoption and usage of GenAI by students, with a primary emphasis on how they integrate and engage with this technology (e.g., Kubullek *et al.*, 2024). So far, the literature on students' GenAI usage has not focused on the intersection between education and business, where part-time university students, who work alongside their studies, are particularly active. Research on part-time students who balance work and study is essential for understanding how GenAI can support their limited time resources, enhance learning efficiency, and promote equal access to academic success despite competing demands. The present study aims to contribute to this scientific discourse by developing a grounded model to understand the GenAI usage of students who navigate between education and business. This leads to the following research question: *Which characteristics designate GenAI usage of university students navigating between education and business?* This research question is addressed using a qualitative research approach, combining the grounded theory method (GTM) and a case study of a distance learning university, where students are working alongside their studies and are using GenAI, especially ChatGPT, for educational and business purposes. Eleven semi-structured interviews were conducted with students of the given university. There are two objectives of this study: First, exploring the characteristics that designate students' GenAI usage in education and business. Second, developing a grounded model of the phenomenon 'GenAI usage'. The results of this study aim to provide deeper insights into the usage of GenAI, particularly by identifying influencing factors and strategies to enhance its application at the intersection of education and business. In the next sections, we outline the theoretical background (2.), our GTM and case selection (3.), our results (4.), and we discuss the findings, outline the limitations of this study and the implications for research and practice (5.).

## 2  Theoretical background

Recent advancements in AI enable the creation of meaningful content—such as text, images, and audio—that is increasingly indistinguishable from human craftsmanship. This capability is powered by GenAI, a term that refers to computational techniques designed to generate such content based on training data (Feuerriegel *et al.*, 2024). The deployment of GenAI has resulted in significant advancements and transformations across various domains, including the field of education (Adiguzel *et al.*, 2023). GenAI can support instructors and students in various subjects, but there are also potential pitfalls, such as the generation of fake information and its impact on academic integrity (Lo, 2023). GenAI can be used for academic misconduct and cheating, especially in cases where students do not perceive the use of AI as cheating (Cotton *et al.*, 2024). The field of plagiarism and GenAI usage, also named as "AI-giarism" (Chan, 2023), describes the blurring of traditional boundaries of authorship and plagiarism, thus raising questions about academic integrity in the age of AI. Current research is predominantly focused on the acceptance and adoption of GenAI by students (e.g. Duong, 2024; Strzelecki, 2024; Shuhaiber *et al.*, 2025). Annamalai *et al.* (2025) stated that the usage of ChatGPT is influenced by autonomy and relatedness, which have been shown to have a positive correlation to the development of competencies in the use of GenAI. This ultimately results in sustained use (Annamalai *et al.*, 2025). Kubullek *et al.* (2024) add with their study that the adoption of GenAI is primarily influenced by the academic discipline with STEM students showing a higher willingness to adopt compared to economics students. However, overall most participants showed a positive attitude towards AI, reflecting its growing acceptance in education (Kubullek *et al.*, 2024). Monib *et al.* (2025) examined the experiences of students with ChatGPT and identified several key benefits of its usage, including language learning support, personalized interactive experiences, and streamlined research processes. However, alongside these advantages, learners have articulated concerns including ChatGPT's accuracy, reliability, up-to-datedness and the challenges involved in verifying the generated content (Monib *et al.*, 2025). In business contexts, GenAI can unleash innovation and streamline work processes (Rajaram & Tinguely, 2024) and increases productivity in different kinds of business departments like marketing, management and development (Walke *et al.*, 2025). Among others, GenAI shows potential to business analytics (Salazar & Kunc, 2025), improves scenario planning (Finkenstadt *et al.*, 2024), and influences organizational performance (Rana *et al.*, 2024).

Previous research has already explored the adoption (Kubullek *et al.*, 2024), acceptance (e.g. Liu & Zhang, 2024), usage (e.g. Wecks *et al.*, 2024; Walke *et al.*, 2025), and the impact (e.g. Rana *et al.*, 2024; Saúde *et al.*, 2024) of GenAI separately in business and education. However, research on the intersection of both domains remains lacking. It is important to note that today's students will eventually become tomorrow's employees (Walstrom & Duffy, 2003), underscoring the potential value of experiences gained during university for smooth transition into the professional world. Additionally, research aiming to gain a more profound understanding of the impact of GenAI on students is still ongoing. A contributing factor is the rapid evolution of the GenAI landscape, with advancements promised in each release (Slowik, 2024). To date, as far known there has been no research undertaken with the objective of developing a grounded model to understand the usage of GenAI by students navigating between education and business, particularly by employing the GTM (Corbin & Strauss, 2015).

## 3 Methodology

The grounded theory methodology was applied, following the paradigm described by Strauss & Corbin (Strauss & Corbin, 1990; Corbin & Strauss, 2015) and the GTM procedure described by Wiesche et al. (2017). This study is examined in the European Union, whereby a distance learning university in Germany serves as the case. Using case data to build grounded theory has the strengths that it is likely to: produce novel theory, be testable with constructs that can be readily measured, be empirically valid (Eisenhardt, 1989). The study pursued a qualitative research approach, by conducting student interviews and analyzing them qualitatively. The study took place over a period of three months in 2024 and 2025. Methodologically, this GTM study was based on eight steps described by Wiesche et al. (2017). During the steps of theoretical sampling (1) and role of prior theory (2), and in addition to identifying the research gap, deriving the research questions and data acquisition, we used Strauss and Corbin's GTM paradigm (Strauss & Corbin, 1990) to define a GTM study model and to design our interview guidelines for the interviews. Our GTM study model (Figure 1) consists of the categories causal conditions, context, strategies, intervening conditions and consequences. For the strategies category, we employ the TIHPS framework (Walke, 2024; Walke & Winkler, 2024) as our methodological foundation. Regarding the connections between the predefined GTM categories, we do not claim to identify causal relationships between the categories. Instead, we have used the relationships between the categories outlined by Corbin & Strauss (2015) as the basis for our interview guide.

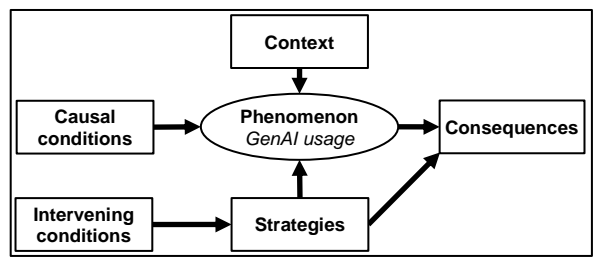

**Figure 1.** GTM study model; adapted from Corbin & Strauss (2015)

The phenomenon was predetermined as generative artificial intelligence usage (GenAI usage). Eleven students were interviewed for the data collection. There were three criteria for participation in the interviews: First, the interviewees had to be enrolled part-time students, second, they work in an enterprise besides their study, and third, the interviewees had to be experienced in using GenAI tools, such as ChatGPT. For the first criterion, we also considered it valid if the majority of their study time is spent with an employer. In the course of the qualitative interview study, ethical principles such as anonymous and confidential treatment of the data, voluntariness and informed consent, as well as protection against impairment and harm to the subjects were taken into account. The interviews were processed, based on the interview guidelines, with open and explorative questions related to the given GTM study model category. The data analysis steps of open coding (3), axial coding (4), selective coding (5), constant comparison (6) and memoing (7) described by Wiesche et al. (2017) have been performed using MAXQDA Software and were based on the coding system of the GTM study model (Figure 1). According to Strauss & Corbin (1990), and in contrast to Wiesche et al. (2017), coding was completed after open, axial and selective coding. The step of the final connection to the coding paradigm (8) was used to develop the final grounded model.

## 4      Results

We present the results of this study according to the categories of the GTM study model, distinguishing between the axial and selective coding steps, as these provide compressed information value compared to open coding. This presentation of the findings is applied to provide a logical chain of evidence. The presentation of the coding results is based on the explanations of Williams & Moser (2019). We highlight relevant findings of the coding process with direct citations from the participants.

**Table 1.** Context and its coding

| Axial Coding | Selective Coding |
|---|---|
| Subject; permission | Application context |
| Personal background; professional background; specialized knowledge | Personal context |

The *application context* contains the axial codes *subject* and *permission*, which refer to the circumstances under which the phenomenon is occurring. In the context of university studies, the participants repeatedly highlighted that the benefit of GenAI depends on the subject: *"So I think what I would definitely say is that it depends very much on the subject I'm studying. Well, there are subjects, (…) where I realized very quickly: No chance, it won't work, the answers aren't right"*. *Permission* indicates that its use for specific tasks is permitted, or not prohibited: *"another condition would of course be that I am allowed to use it in that context. So, for example, I now logically know that I am not allowed to have entire texts written with it in order to then publish them"*. *Permission* is not only relevant for study-related tasks but also in the professional business environment. One participant shared the insight that their company had blocked all GenAIs: *"I said that I work at the bank and we have super-sensitive customer data, which is why we're not allowed to use one per se, so that there's no risk of any sensitive data being inserted"*. In *personal context*, the *personal background*, *professional background* and *specialized knowledge* have been identified. *Personal background* was mainly associated with age. The participant noted that GenAI usage is more common in younger age groups. Also, the *professional background*, which is defined by the work environment someone is working in, was mentioned: *"the area in which you work. I think that does play an important role"*. Regarding *specialized knowledge*, the lack of this knowledge was mentioned as a contextual code: *"But I don't think many people know how they can create value from this."*

The term *limiting conditions* is applied to any factor that restricts the functionality of GenAI tools and thus, impacting its usage. The *limitations of GenAI* in academia refer to scenarios where GenAI, such as ChatGPT, do not produce satisfactory results, particularly in multiple-choice tasks, for example. *System resources* describe instances where ChatGPT is overloaded or freezing. *System limitations* refer to several weak-nesses for GenAI systems, such as the production of erroneous results and hallucinations, the presence of bias in GenAI-generated output, and the reliability of the outputs. Other limitations are related to system features, such as problems when big documents are uploaded to ChatGPT.

Table 2. Causal conditions and their coding

| Axial Coding | Selective Coding |
| --- | --- |
| Limitations of GenAI; system resources; system limitations; | Limiting conditions |
| Personal experience; curiosity; fun; success; social acceptance; | Personal conditions |
| System applicability; suitable use cases; task complexity; advantages of ChatGPT; | Procedural conditions |

Overall, all participants demonstrated a good understanding of these limitations, as illustrated by the following statement by one participant: *"You can't blindly trust everything he says. If you are not yet familiar with a topic, it makes little sense to ask ChatGPT, because then you simply can't judge: Does it make sense or not?"* The production of incorrect or hallucinated output was identified as the most significant constraint by all participants. Many participants reported that the most frustrating situations occurred when a conversation stagnated, as illustrated by the following statement: *"And the second huge problem is that he might be going round in circles. So, he never says: I have no idea. Instead, he gives the same and the same and the same answer over and over again. Sometimes I even got a bit angry with him"*. Another drawback is that the user must always define the context for the interaction with GenAI, as illustrated by this participant: *"It's always like a newborn and you always have to introduce it to what you want, what the status quo is"*.

*Personal conditions* are defined as user-related factors that influence the usage of GenAI, including *personal experience*, *curiosity*, *fun*, *success*, and *social acceptance*. Among these factors, *curiosity* emerges as a leading driver of GenAI usage. The *procedural conditions* consist of *system applicability*, *suitable use cases*, *task complexity*, and *advantages of ChatGPT*. The usage of GenAI is dependent upon *suitable use cases*, which, once identified, can facilitate the delivery of solutions for *complex tasks*, thereby fostering a positive experience and, in turn, promoting the usage of GenAI. *System applicability* signifies the broad spectrum of applicability of GenAI: *"I think it's super valuable because you can really do a lot with it."* Consequently, expectations are notably high: *"I also have to say that I now expect it to be able to do quite a lot"*. The final component of the procedural conditions addresses the *advantages of ChatGPT*, a subject that was addressed by the majority of participants. However, other GenAI tools, such as Claude, Perplexity and Copilot, were also mentioned. Overall, the participants expressed that ChatGPT offers the most comprehensive user experience. Another notable advantage of ChatGPT over its competitors is the absence of advertisements on its webpage.

The selective code *user trust dynamics* contains *attitude and perception of the user*. While the participants showed a positive attitude towards GenAI, they also mentioned that GenAI is hyped like this participant articulates: *"I think that's the hype that comes*

*to mind, because I think you can attribute the fact that AI has attracted so much attention recently to it."* However, they also noticed that GenAI is changing different aspects, for example, the way we do tasks, the way we think, and work, which results in the fact that GenAI is changing jobs. The participants' trust in the outcomes generated by GenAI ranges from low to high, with higher trust levels observed for easier tasks and lower levels for more complex tasks. Two axial codes were identified for the *user qualification*: *competence acquisition* and *user competence requirements*. It was noted that users must understand the nature of GenAI, including its capabilities and limitations, to make informed decisions regarding its application. This concept is closely related to the *specialized knowledge* identified in *context*. The incorporation of this knowledge has the potential to enhance the usage of GenAI. Additionally, the concept of *competence acquisition* refers to the expectation among participants that proficiency in GenAI will become a mandatory component of professional development in the future like this participant is describing: *"I also think it prepares you for the working world, because everyone will be working with ChatGPT or GenAI in the future"*. Through the integration of GenAI into their academic endeavors, participants can prepare themselves for potential professional requirements.

**Table 3.** Intervening conditions and their coding

| Axial Coding | Selective Coding |
|---|---|
| Costs; GenAI system design; quality of output; | GenAI conditions |
| Interaction requirements; user experience; | User interaction |
| Competence acquisition; user competence requirements; | User qualification |
| Attitude and perceptions of the user; | User trust dynamics |

The concept of *user interaction* involves two axial codes: *interaction requirements* and *user experience*. A component of these requirements concerns the accessibility of GenAI tools and their user-friendliness. The participants noted that the complexity of account creation hinders usage, while easy access modalities like ChatGPT can promote increased usage. The seamless and intuitive interface of ChatGPT shows a positive *user experience* that fosters continuous usage. Another crucial aspect addresses the interaction with the system and the speed at which users get satisfactory results. When users experience efficient information access, it leads to an increase in usage. This behavior contributes to the overall *user experience*. The experience users have when using GenAI has an immediate effect on usage, and a positive experience, such as faster learning success or the confidence to ask "dumb" questions without fear of being judged, resulting in the ability to solve complex tasks, has an immense increasing effect on the use of GenAI. Conversely, a poor experience with GenAI decreases the willingness to use this GenAI and impacts the willingness to try new GenAI tools. The intervening condition *GenAI conditions* contains: *costs*, *GenAI system design*, and *quality of output*. The factors *costs* and the *quality of the output* were identified as major influence factors and thus separated. When using GenAI tools, functionality is ultimately limited when the user does not pay for it. For instance, the free version of ChatGPT offers access to specific models and imposes limitations on the number of prompts users can utilize each day. These limitations are often removed through paid subscriptions, which can significantly impact usage rates. Similarly, the quality of the output generated has comparable effects. Ensuring the accuracy and reliability of the output is the most frequently expressed expectation when using GenAI: *"In other words, it's really important to me that the answers are good or good enough that I can use them and that they're not too wrong. Otherwise, I can google it too"*. Other determining factors include the increasing functionality and *quality of outputs*, as well as its growing integration into various systems, such as development environments as explained by this participant: *"How it is integrated into the tools I use. I already notice that when I open the database, the database program, which doesn't have ChatGPT, it annoys me that I have to copy things out"*. Additionally, the protection of data privacy of data handled by GenAI systems is a crucial consideration, particularly when dealing with sensitive information. Conversely, poorly designed GenAI has been identified as a relevant weakening factor.

**Table 4.** Strategies and their coding

| Dimension | Axial Coding | Selective C. |
|---|---|---|
| Technology | Increase availability, integration, interactivity, navigability, quickness, sovereignty; | GenAI improvement |
| Information | Increase accuracy, findability, relevance, output reliability, data protection; | |
| Process | Increase transparency; | |
| System | Adapt regulations; increase security, resources, sustainability; | |
| | GenAI training provision; learning from existing use cases; | Communication and Training |
| Human | Input related strategies; output related strategies; personal strategies; user awareness strategies; | Usage adaptation |

The highest-mentioned strategy regarding *GenAI improvement* focuses on *security* and *data protection (adapt regulations)*. The participants are keenly aware that, for instance, the inputs to ChatGPT are used to train the model. Consequently, they emphasized the importance of carefully formulating inputs to prevent the leakage of sensitive information, particularly in professional contexts where users act with particular caution. In addition, participants expressed a wish for enhanced data governance and the capability to remove information, such as the option to delete an account. Another noticeable point is the expectation of optimized *resource* consumption and increased *sustainability* of these tools, given their current high energy consumption. The *output reliability* was identified as an important factor influencing its usage. Consequently, expectations regarding *GenAI adaptations* and improvements in this area were among the most mentioned. These improvements are expected to increase the accuracy of the output and promote *transparency* regarding *output reliability*, which consistently generate outputs irrespective of how reliable they are: *"It would be cool if he could underline how sure he is about the information he gives me"*. With regards to *interactivity*, the most frequently requested improvement is related to two key aspects. First, there is the demand for additional interaction modes, such as talking to GenAI or engaging with drawings. Second, there is a demand for improvements in the output generation process. Participants have noted that when a prompt is entered, GenAI becomes occupied with processing it, preventing any interaction until this processing is complete. To enhance the efficiency of interaction, participants proposed the capability to discontinue the processing of a prompt and the option to reduce the output length as described by this participant. Additionally, as previously mentioned, the absence of context was identified as a factor influencing the usage of GenAI. Consequently, enhancements in the context interpretation of GenAI tools were proposed as a solution.

The integration of GenAI in various situations requires a range of strategies by users. These strategies can be categorized into four distinct categories: *input-related*, *output-related*, *personal* and *user awareness strategies*. Among the *input-related strategies*, prompt engineering, repeated questioning, and initiating a new request when facing difficulties are particularly prominent in improving the efficacy of GenAI usage. *Output-related strategies* primarily center on verifying the outcome, e.g., by conducting a web search for the result, and providing feedback to the output (e.g., in ChatGPT, an easy feedback option is available with a simple thumbs-up or thumbs-down voting). *Personal strategies* refer to procedures by users to increase the benefits resulting from GenAI usage. These include the delegation of specific tasks to GenAI to minimize mental load, the adoption of skills to broaden one's GenAI knowledge, and the trying of novel prompts. Basic aspects of *personal strategies* include the evaluation of outcomes through common sense and, in some cases, the decision to discontinue. Additionally, the relationship between the user and GenAI has been identified as crucial, particularly regarding the aspect of missing context. This is illustrated by the observation that as GenAI systems become more familiar with a user, the amount of context required for specific tasks reduces. The strategy *user awareness* primarily centers on the necessity of handling GenAI with caution, particularly with sensitive data. For instance, it has been observed that participants possess a clear understanding of the appropriateness of inserting internal company information only in company-internal GenAI solutions, as demonstrated by this participant: *"So I think it's good that the company has its own company GPT as a replacement for ChatGPT because we're not allowed to use ChatGPT for anything close to internal data or anything like that, which is a good thing"*. Communication and Training, contains *GenAI training provision* and *learning from existing use cases*. Participants mentioned that they had previously joined GenAI trainings, with some of these trainings being offered by their university. However, they stated that they would prefer to have more training opportunities, particularly within academia. Also, the participants shared personal experience with academic tasks they had encountered, in which GenAI usage played a pivotal role. Positive examples could serve as a model for other universities or departments to incorporate GenAI more seamlessly into their academic curricula, thus learning from *existing use cases*.

The *consequences* of GenAI usage show a notable increase in productivity. This *productivity increasement* results in increased *study efficiency*, *work effectiveness* and overall *optimized performance*, leading to *time savings*. The participants mostly reported using GenAI as a support tool for writing tasks, such as translation and adaptation of language, while expressing a strong aversion to using GenAI-generated content for graded academic work. The second most mentioned use case is the usage of GenAI to obtain inspiration or new ideas on a topic, followed by the use cases of exam preparation and literature research. In general, the participants explained that they are more productive because GenAI takes on non-value-added work, which is why they continue to use GenAI for tasks, especially in the academic context (e.g., summarizing content, debugging codes, preparing slides). The most relevant practical consequence of GenAI usage is that it seems to replace classical internet search tools, such as Google. Nearly all participants reported using GenAI instead of Google, because it provides the requested information more quickly. They noted that using GenAI is faster than scrolling through and evaluating search results to identify the information they seek, as described by this participant: *"Sometimes I use it because I'm just too lazy to google. Googling is a skill itself"*. Other effects can be classified as either negative or positive. The negative *effects* that were mentioned include excessive reliance on the outcomes generated by GenAI, the usage of GenAI not resulting in high-quality results, and the taking of shortcuts that ultimately leads to laziness. Conversely, the positive *effects* that were mentioned included the fact that GenAI can compensate for disadvantages and bridge communication gaps with experts. *Ethical implications* arise from the usage of GenAI, particularly in the context of participants expressing concerns about the level of GenAI usage that is appropriate and when GenAI usage becomes cheating. In a more holistic sense, the question of GenAI's role in the future and the extent to which it should emulate human characteristics arose. Nevertheless, the *perception and evaluation* was influenced by their general *satisfaction* with GenAI, although several instances of *dissatisfaction* were also described.

**Table 5.** Consequences and their coding

| Axial Coding | Selective Coding |
|---|---|
| Future implications of GenAI; GenAI training courses | Future significance |
| New competencies; occurring learning effects; replacing skills and knowledge; | Learning promotion |
| Limits of GenAI usage; risks of GenAI usage; | Limitations and risks |
| Satisfaction with GenAI; dissatisfaction with GenAI; ethical implication; effects of GenAI; | Perception and evaluation |
| Increasing work effectiveness; increasing study efficiency; optimized performance; saving of time | Productivity increasement |

A potential *risk* was identified: the possibility of falsely being accused of plagiarism and cheating when utilizing GenAI tools, particularly when the guidelines provided by universities are not clear. This concern is expressed by a participant as follows: *"I'm extremely anxious about using AIs in my studies because I'm afraid that I could be accused of plagiarism, for example in some seminar papers or even now in my Master's thesis, and I think that's a real shame because you use Google or even your parents to correct your work and correcting wording is absolutely allowed"*. The *limitations* of GenAI usage are based on the limiting conditions, particularly the hallucinations of GenAI systems. As the generated output may contain imaginary facts, academic literature remains the primary source for knowledge. Nevertheless, the usage of GenAI has been proven to *promote learning*, particularly by facilitating engagement with topics, thereby adding value to the educational journey. Additionally, GenAI has been observed to foster resilience, encouraging individuals to continue their efforts without surrender (*occurring learning effects*). The application of GenAI requires the development of new *competencies*, such as prompt engineering, and serves as an entry point into the broader field of artificial intelligence. However, it is important to note that GenAI has the potential to *replace certain skills and knowledge* areas, such as coding. One participant mentioned that they received advice to prioritize learning to create prompts rather than learning a programming language: *"I am now often told that I should not invest the time in practicing coding myself but rather invest the time in formulating good prompts and practicing prompting"*. The *future significance* of GenAI, in which nearly all participants agreed that it will become increasingly significant. This anticipation is further explained by the anticipation that employees who are not familiar with GenAI tools may face disadvantages in the future, increasing the demand for *GenAI training courses*. Additionally, there was a concern that GenAI could potentially replace jobs (*future implications*). In the academic context, there was a fear that university degrees could lose value due to GenAI. One potential effect of this transformation is a paradigm shift in the requirements for academic qualification, such as the necessity of writing a thesis. Based on the underlying study model (Figure 1), the findings of this study are assembled as a grounded model of students' GenAI usage in Figure 2.

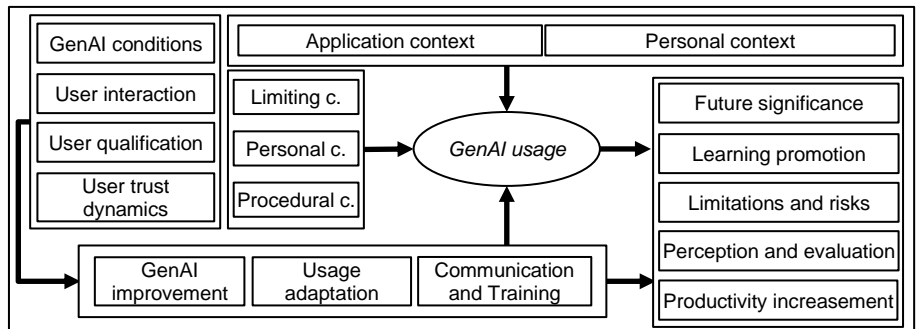

**Figure 2.** Grounded model of university students' GenAI usage

## 5 Discussion and Conclusion

The research question "*Which characteristics designate GenAI usage of university students navigating between education and business?*" was addressed through a qualitative research approach based on the grounded theory methodology, which resulted in a grounded model (Figure 2). This study contributes to the scientific debate by providing a model which describes the usage of GenAI by students in education and business, thereby encompassing a comprehensive and diverse perspective. In general, through the academic connection the students of this study seem to have a broader and more socially conscious perspective on GenAI-related strategies, conditions and consequences, compared to full-time enterprise employees, who are more focused on department-specific tasks (Walke *et al.*, 2025), as their academic connection has diminished. This study is in line with several categories of other GenAI usage models (Walke *et al.*, 2025) and other previous research results. For example, GenAI seem to be generally used by students when they experience an immediate benefit in their academic endeavors, which is in line with Strzelecki (2024). The student's user experiences were identified as a major intervening condition. Thus, improving GenAI user experiences can have an impact on GenAI usage, which is in line with the results from Liu and Zhang (2024). Additionally, this study shows that students have high expectations on GenAI systems, a finding that was also described by Chan and Hu (2023). Also the findings of Monib *et al.* (2025) regarding concerns and strategies matches the findings from this study. Chan and Hu (2023) show that there is no significant correlation between students' concerns and their knowledge of GenAI, indicating that even well-informed students may still have reservations. This observation is consistent with the results of this study. Although the students might have concerns and a lower trust in GenAI generated outcome, they showed an overall positive attitude towards GenAI, a finding also described by Walke et al. (2025) and Kubullek *et al.* (2024). Some students encountered academic tasks where the use of GenAI was crucial. These tasks, in turn, helped them understand both the strengths and limitations of GenAI. In general, the students expressed their wish to include GenAI in academic tasks. Annamalai *et al.* (2025) and Kubullek *et al.* (2024) formulated similar requests. The same applies for general training on GenAI which was requested by the participants of this study. Here, also Duong (2024) expressed that educational institutions should invest in comprehensive training programs and readily accessible technical support.

A theoretical contribution of this study is, that the TIHPS framework (Walke, 2024; Walke & Winkler, 2024) proves to be applicable for identifying and evaluating improvement and usage strategies related to GenAI. Here, additional quality indicators were identified: integration and availability for technology quality; veracity for the information quality and relationship for the human quality factor. These results could be used as a basis for further research to adapt and evaluate frameworks to evaluate the quality of GenAI tools. A practical implication is, that the expressed concerns, especially around academic integrity, should be addressed by educators, also the request for guidelines and regulations should be taken up by universities and other educational institutions. The results of this study also show that GenAI should not be banned in general. Instead, its strengths and weaknesses should be made transparent through targeted integration into the academic curriculum. With the anticipated significance of GenAI in the future, educational institutions must prepare students for their professional lives by equipping them with GenAI knowledge and the necessary GenAI skills to succeed.

This study only utilizes a small sample size of 11 participants, all located in Germany. Although theoretical saturation was reached, the validity of the model would benefit from additional and more diverse participants in future research. Our implications can be restricted by our sample. In this study, a grounded model was developed to describe the use of GenAI by students who simultaneously work and study. Since all participants have work experience, the model also captures aspects of GenAI usage beyond academic endeavors. The findings can contribute to a more comprehensive understanding of GenAI usage, supporting the development of effective educational applications and guidelines for designing and improving GenAI technology for both educational and business-related applications.